\documentstyle[aps,psfig]{revtex}

\newcommand{\be}{\begin{equation}}
\newcommand{\ee}{\end{equation}}
\newcommand{\ben}{\begin{eqnarray}}
\newcommand{\een}{\end{eqnarray}}

\begin{document}

\twocolumn[\hsize\textwidth\columnwidth\hsize\csname 
@twocolumnfalse\endcsname 
 
\title{Superconducting transition temperature in thin films} 
\author{A. P. C. Malbouisson$^\star$, J. M. C. Malbouisson$^\dagger$,
and A. E. Santana$^\dagger$}
\address{$^\star$Centro Brasileiro de Pesquisas F\'\i sicas, Rua Dr.
Xavier Sigaud 150, Urca,
22290-180, Rio de Janeiro, RJ, Brazil\\
$^\dagger$Instituto de F\'\i sica, Universidade Federal da Bahia, Campus  
de Ondina, 40210-340, Salvador, BA, Brazil} 
\date{\today} 
 
\maketitle 
 
\begin{abstract}
By considering the Ginzburg-Landau model, compactified in one of the spatial 
dimensions, and using a modified Matsubara formalism, we determine the
dependence of the superconducting transition temperature ($T_{c}$) of a film
as a function of its thickness ($L$). We show that $T_{c}$ is a decreasing
linear function of $L^{-1}$, as has been found experimentally. The critical
thickness for the suppression of superconductivity is expressed in terms
of the Ginzburg-Landau parameters.
\\
\\
{PACS number(s): 74.20.De, 11.10.-z, 05.70.Fh, 74.76.-w}\\
\end{abstract} 
\vskip2pc]

\newpage

In last decades, a large amount of work has been done on the Ginzburg-Landau
model applied to the study of the superconducting transition, both in the
single component and in the $N$- component versions of the model, using the
renormalization group approach. The state of the subject, for type-I and
type-II superconductors and related topics, can be found for instance in
Refs. \cite{Afleck,Lawrie,Lawrie1,Brezin,Radz,Flavio1}. In another related
topic of investigation, there are systems that present domain walls as
defects, created for instance in the process of crystal growth by some
prepared circumstances. At the level of effective field theories, in many
cases, this can be modeled by considering a Dirac fermionic field whose mass
changes sign as it crosses the defect, meaning that the domain wall plays
the role of a critical boundary separating two different states of the
system \cite{Cesar,FAdolfo}. Questions concerning stability and the
existence of phase transitions may also be raised if one considers the
behavior of field theories as function of spacial boundaries. Studies on
confined field theory have been done in the literature since a long time
ago. In particular, an analysis of the renormalization group in finite size
geometries can be found in Refs. \cite{Zinn,Cardy}. These studies have been
performed to take into account boundary effects on scaling laws. The
existence of phase transitions, would be in this case associated to some
spatial parameters describing the breaking of translational invariance, for
instance the distance $L$ between planes confining the system. In this
situation, for Euclidean field theories the Matsubara formalism applies for
the breaking of invariance along any one of the spacial directions. Studies
of this type have been recently performed \cite{JMario,Ademir}, concerning
with the spontaneous symmetry breaking in the $\lambda \phi ^{4}$ theory. In
particular, if one considers the Ginzburg-Landau model confined between two
parallel planes, thus describing a superconducting film, the question of how
the critical temperature depends on the thickness $L$ of the film can be
raised.

Under the assumption that information about general features of the behavior
of superconductors, in absence of magnetic fields, can be obtained in the
approximation which neglects gauge field contributions in the
Ginzburg-Landau model, in this letter we examine this model with an approach
different from the renormalization group analysis. We consider the system
confined between two parallel planes and we use the formalism developed in
Refs. \cite{JMario,Ademir} to investigate how the critical temperature is
affected by the presence of boundaries. From a physical point of view, we
investigate how the critical temperature of a superconducting film depends
on its thickness.

We start with the Ginzburg-Landau Hamiltonian density in the Euclidean $D$
-dimensional space, in absence of magnetic fields, given by (in units with $ 
\hbar =1$) 
\begin{equation}
{\cal H}=\left| {\bf \nabla }\varphi \right| ^{2}+m_{0}^{2}\left| \varphi
\right| ^{2}+\frac{\lambda }{2}\left| \varphi \right| ^{4}\,,  \label{H}
\end{equation}
where $\lambda $ is the (renormalized) self-coupling constant, with the
``bare mass'' given by $m_{0}^{2}=\alpha (T-T_{0})$, $T_{0}$ being the bulk
transition temperature of the superconductor and $\alpha >0$. We consider
the system confined between two parallel planes, normal to the $x$-axis, a
distance $L$ apart from one another and use Cartesian coordinates ${\bf r}
=(x,{\bf z})$, where ${\bf z}$ is a $(D-1)$-dimensional vector, with
corresponding momenta ${\bf k}=(k_{x},{\bf q})$, ${\bf q}$ being a $(D-1)$
-dimensional vector in momenta space. The partition function is written as, 
\begin{equation}
{\cal Z}=\int {\cal D}\varphi ^{\ast }{\cal D}\varphi \exp \left(
-\int_{0}^{L}dx\int d^{D-1}{\bf z}\;{\cal H}(|\varphi |,|\nabla \varphi
|\right) \,,  \label{Z}
\end{equation}
with the field $\varphi (x,{\bf z})$ satisfying the condition of confinement
along the $x$-axis, $\varphi (x\leq 0,{\bf z})\;=\;\varphi (x\geq L,{\bf z}
)\;=\;0$. Then the field should have a mixed series-integral Fourier
representation of the form, 
\begin{equation}
\varphi (x,{\bf z})=\sum_{n=-\infty }^{\infty }c_{n}\int d^{D-1}{\bf q}\;b( 
{\bf q})e^{-i\omega _{n}x\;-i{\bf q}\cdot {\bf z}}\tilde{\varphi}(\omega
_{n},{\bf q})\,,  \label{Fourier}
\end{equation}
where $\omega _{n}=2\pi n/L$ and the coefficients $c_{n}$ and $b({\bf q})$
correspond respectively to the Fourier series representation over $x$ and to
the Fourier integral representation over the $(D-1)$-dimensional ${\bf z}$
-space. The above conditions of confinement of the $x$-dependence of the
field to a segment of length $L$ allow us to proceed, with respect to the $x$
-coordinate, in a manner analogous as it is done in the imaginary-time
Matsubara formalism in field theory and, accordingly, the Feynman rules
should be modified following the prescription 
\begin{equation}
\int \frac{dk_{x}}{2\pi }\rightarrow \frac{1}{L}\sum_{n=-\infty }^{+\infty
}\;,\;\;\;\;\;\;k_{x}\rightarrow \frac{2n\pi }{L}\equiv \omega _{n}\,.
\label{Matsubara}
\end{equation}
We emphasize that we are considering an Euclidean field theory in $D$ {\it  
purely} spatial dimensions, so we are {\it not} working in the framework of
finite temperature field theory. Here, the temperature is introduced in the
mass term of the Hamiltonian by means of the usual Ginzburg-Landau recipe.

To continue, we use some one-loop results described in \cite{JMario,Gino1},
adapted to our present situation. These results have been obtained by the
concurrent use of dimensional and $zeta$-function analytic regularizations,
to evaluate formally the integral over the continuous momenta and the
summation over the Matsubara frequencies. We get sums of polar ($L$
-independent) terms plus $L$-dependent analytic corrections. Renormalized
quantities are obtained by subtraction of the divergent (polar) terms
appearing in the quantities obtained by application of the modified Feynman
rules (Matsubara prescription) and dimensional regularization formulas.
These polar terms are proportional to $\Gamma $-functions having the
dimension $D$ in the argument and correspond to the introduction of
counter-terms in the original Hamiltonian density. In order to have a
coherent procedure in any dimension, these subtractions should be performed
even for those values of the dimension $D$ where no poles of $\Gamma $
-functions are present. In these cases a finite renormalization is performed.

In the following, to deal with dimensionless quantities in the
regularization procedure, we introduce parameters $c^{2}=m^{2}/4\pi ^{2}\mu
^{2}$, $a=(L\mu )^{-2}$, $g=3\lambda /8\pi ^{2}$ and $\phi _{0}=\varphi
_{0}/\mu $, where $\varphi _{0}$ is the normalized vacuum expectation value
of the field (the classical field) and $\mu $ is a mass scale. In terms of
these parameters, the one-loop contribution to effective potential, adapted
to the situation under study, is given by the well known expression \cite{IZ}
\begin{equation}
U_{1}(\phi ,L=\infty )=\mu ^{D}\sum_{s=1}^{\infty }\frac{(-1)^{s+1}}{2s}
g^{s}\left| \phi _{0}\right| ^{2s}\int \frac{d^{D}k}{(k^{2}+c^{2})^{s}}\,,
\label{potefet0}
\end{equation}
where $m$ (entering in $c$) is the renormalized mass for $L=\infty $.
Performing the Matsubara replacement (\ref{Matsubara}), the
boundary-dependent ($L$-dependent) one-loop contribution to the effective
potential can be written in the form 
\begin{eqnarray}
U_{1}(\phi ,L) &=&\mu ^{D}\sqrt{a}\sum_{s=1}^{\infty }\frac{(-1)^{s}}{2s}
g^{s}\phi _{0}^{2s}  \nonumber \\
&&\times \sum_{n=-\infty }^{+\infty }\int \frac{d^{D-1}k}{(an^{2}+c^{2}+{\bf  
k}^{2})^{s}}\,.  \label{potefet1}
\end{eqnarray}
Now, using the dimensional regularization formula, 
\begin{equation}
\int \frac{d^{d}k}{(2\pi )^{d}}\frac{1}{\left( k^{2}+M\right) ^{s}}=\frac{
\Gamma \left( s-\frac{d}{2}\right) }{(4\pi )^{\frac{d}{2}}\Gamma (s)}\frac{1 
}{M^{s-\frac{d}{2}}}\,,  \label{dimreg}
\end{equation}
Eq. (\ref{potefet1}) reduces to

\begin{equation}
U_{1}(\phi ,L)=\mu ^{D}\sqrt{a}\sum_{s=1}^{\infty }f(D,s)g^{s}\phi
_{0}^{2s}Z_{1}^{c^{2}}(s-\frac{D-1}{2};a)\,,  \label{potefet2}
\end{equation}
where $f(D,s)$ is a function proportional to $\Gamma (s-\frac{D-1}{2})$ and $ 
Z_{1}^{c^{2}}(s-\frac{D-1}{2};a)$ is one of the Epstein-Hurwitz $zeta$
-functions, defined by 
\begin{equation}
Z_{K}^{c^{2}}(u;\{a\})=\sum_{n_{1},...,n_{K}=-\infty }^{+\infty }\frac{1}{
(a_{1}n_{1}^{2}+...+a_{K}n_{K}^{2}+c^{2})^{u}},  \label{zeta}
\end{equation}
valid for $Re(u)>\;K/2$ (in our case $Re(s)>\;D/2$).

The Epstein-Hurwitz $zeta$-function can be extended to the whole complex $s$
-plane and we obtain, after some manipulations \cite{Elizalde}, the one-loop
correction to the effective potential, 
\begin{eqnarray}
U_{1}(D,L) &=&\mu ^{D}\sum_{s=1}^{\infty }g^{s}\phi _{0}^{2s}h(D,s) 
\nonumber  \label{potefet3} \\
&&\times \left[ 2^{-(\frac{D}{2}-s+2)}\Gamma (s-\frac{D}{2})(m/\mu
)^{D-2s}+\right.  \nonumber \\
&&\left. +\sum_{n=1}^{\infty }(\frac{m}{\mu ^{2}nL})^{\frac{D}{2}-s}K_{\frac{
D}{2}-s}(mnL)\right] ,
\end{eqnarray}
where 
\begin{equation}
h(d,S)=\frac{1}{2^{D/2-s-1}\pi ^{D/2-2s}}\frac{(-1)^{s+1}}{s\Gamma (s)}
\label{h}
\end{equation}
and $K_{\nu }$ are the Bessel functions of the third kind.

Note that since we are using dimensional regularization techniques, there is
implicit in the above formulas a factor $\mu ^{4-D}$ in the definition of
the coupling constant. In what follows we make explicit this factor, the
symbol $\lambda $ standing for the dimensionless coupling parameter (which
coincides with the physical coupling constant in $D=4$). We work in the
approximation of neglecting the $L$-dependence of the coupling constant,
that is we take $\lambda $ as the {\it renormalized} coupling constant. In
this case, it is enough for us to use only one renormalization condition, 
\begin{equation}
\left. \frac{\partial ^{2}}{\partial \phi ^{2}}U_{1}(D,L)\right| _{\phi
_{0}=0}=m^{2}\mu ^{2}.  \label{renorm1}
\end{equation}
Since we are using a modified minimal subtraction scheme, where the mass
(and coupling constant, if it is the case) counter-terms are poles at the
physical values of $s$, the $L$-dependent correction to the mass is
proportional to the regular part of the analytical extension of the
Epstein-Hurwitz $zeta$-function in the neighborhood of the pole at $s=1$.
Thus the $L$-dependent renormalized mass, at one-loop approximation, is
given by 
\begin{equation}
m^{2}(L)=m_{0}^{2}+\frac{3\lambda \mu ^{4-D}}{2(2\pi )^{D/2}}
\sum_{n=1}^{\infty }\left[ \frac{m}{nL}\right] ^{(D-2)/2}K_{\frac{D-2}{2}
}(nLm).  \label{massaR}
\end{equation}

On the other hand, if we start in the ordered phase, the model exhibits
spontaneous symmetry breaking, but for sufficiently small values of $T^{-1}$
and $L$ the symmetry is restored. We can define the critical curve $ 
C(T_{c},L_{c})$ as the curve in the $T\times L$ plane for which the inverse
squared correlation length , $\xi ^{-2}(T,L,\varphi _{0})$, vanishes in the $ 
L$-dependent gap equation \cite{Zinn}, 
\begin{eqnarray}
&&\left. \xi ^{-2}=m_{0}^{2}+6\lambda \mu ^{4-D}{\bf \varphi }
_{0}^{2}+\right.  \nonumber \\
&&\;\;\;+\frac{6\lambda \mu ^{4-D}}{L}\sum_{n=-\infty }^{\infty }\int \frac{
d^{D-1}k}{(2\pi )^{D-1}}\;\frac{1}{{\bf k}^{2}+\omega _{n}^{2}+\xi ^{-2}}\,,
\label{gap}
\end{eqnarray}
where ${\bf \varphi }_{0}$ is the normalized vacuum expectation value of the
field (different from zero in the ordered phase). In the disordered phase,
in particular in the neighborhood of the critical curve, $\varphi _{0}$
vanishes and the gap equation reduces to a $L$-dependent Dyson-Schwinger
equation, 
\begin{eqnarray}
m^{2}(T,L) &=&m_{0}^{2}(T)+\frac{6\lambda \mu ^{4-D}}{L}  \nonumber \\
&&\times \sum_{n=-\infty }^{\infty }\int \frac{d^{D-1}k}{(2\pi )^{D-1}}\; 
\frac{1}{{\bf k}^{2}+\omega _{n}^{2}+m^{2}(T,L)}.  \nonumber \\
&&  \label{gap1}
\end{eqnarray}

After steps analogous to those leading from Eq.(\ref{potefet1}) to Eq.(\ref
{massaR}), Eq.(\ref{gap1}) can be written in the form 
\begin{eqnarray}
m^{2}(T,L) &=&m_{0}^{2}(T)+\frac{3\lambda \mu ^{4-D}}{2(2\pi )^{D/2}} 
\nonumber \\
&&\times \sum_{n=1}^{\infty }\left[ \frac{m(T,L)}{nL}\right] ^{(D-2)/2}K_{ 
\frac{D-2}{2}}(nLm(T,L))\,.  \nonumber \\
&&  \label{massaR1}
\end{eqnarray}
If we limit ourselves to the neighborhood of criticality, $m^{2}(T,L)\approx
0$, we may investigate the behavior of the system by using in Eq.(\ref
{massaR1}) an asymptotic formula for small values of the argument of Bessel
functions, 
\begin{equation}
K_{\nu }(z)\approx \frac{1}{2}\Gamma (\nu )\left( \frac{z}{2}\right) ^{-\nu
}\;\;\;(z\sim 0),  \label{K}
\end{equation}
which allows after some straightforward manipulations, to write Eq.(\ref
{massaR1}) in the form 
\begin{equation}
m^{2}(T,L)\approx m_{0}^{2}(T)+\frac{3\lambda \mu ^{4-D}}{(2\pi )^{D/2}}
\Gamma (\frac{D}{2}-1)L^{2-D}\zeta (D-2),  \label{mDysoncr}
\end{equation}
where $\zeta (D-2)$ is the Riemann $zeta$-function, $\zeta
(D-2)=\sum_{n=1}^{\infty }(1/n^{D-2})$, defined for $D>3$. Taking $ 
m^{2}(T,L)=0$ and $m_{0}^{2}=\alpha (T-T_{0})$ in Eq.(\ref{mDysoncr}), we
obtain the critical curve in the $T\times L$ plane for Euclidean space
dimension $D$ ($D>3$), 
\begin{equation}
\alpha (T_{c}-T_{0})+\frac{3\lambda \mu ^{4-D}}{(2\pi )^{D/2}}\Gamma (\frac{
D }{2}-1)L_{c}^{2-D}\zeta (D-2)=0.  \label{mDysoncr1}
\end{equation}

For $D=3$ the Riemann $zeta$-function in Eq.(\ref{mDysoncr1}) has a pole. We
can not obtain a critical curve in dimension $D\leq 3$ by a limiting
procedure from Eq.(\ref{mDysoncr1}). For $D=3$, which corresponds to the
physically interesting situation of the system confined between two parallel
planes embedded in a $3$-dimensional Euclidean space, Eq.(\ref{mDysoncr1})
becomes meaningless. To obtain a critical curve in $D\leq 3$, we perform an
analytic continuation of the $zeta$-function $\zeta (z)$ to values of the
argument $z\leq 1$, by means of the reflection property of $zeta$-functions 
\begin{equation}
\zeta (z)=\frac{1}{\Gamma (z/2)}\Gamma (\frac{1-z}{2})\pi ^{z-\frac{1}{2}
}\zeta (1-z),  \label{extensao}
\end{equation}
which defines a meromorphic function having only one simple pole at $z=1$.
For $D=3$, a mass renormalization procedure can be done as follows:
remembering the formula 
\begin{equation}
\lim_{z\rightarrow 1}\left[ \zeta (z)-\frac{1}{1-z}\right] =\gamma ,
\label{extensao1}
\end{equation}
where $\gamma \cong 0.57216$ is the Euler constant, we define the {\it  
renormalized} mass $\bar{m}$ as 
\begin{eqnarray}
\bar{m}^{2}(T,L) &=&\lim_{D\rightarrow 3_{-}}\left[ m^{2}(T,L)-\frac{
3\lambda \mu }{2\pi \sqrt{2}L(3-D)}\right]  \nonumber \\
&=&\alpha (T-T_{0})+\frac{3\gamma \lambda \mu }{2\sqrt{2}\pi L}\,.
\label{massaRR}
\end{eqnarray}
Taking this {\it renormalized} mass equal to zero leads to the critical
curve in dimension $D=3$, given by 
\begin{equation}
T_{c}=T_{0}-\frac{3\gamma \lambda \mu }{2\sqrt{2}\pi \alpha L_{c}}\,.
\label{critica}
\end{equation}

In Eq.(\ref{critica}), $T_{0}$ corresponds to the transition temperature for
the material in absence of boundaries ($L_{c}\rightarrow \infty $), that is,
to the bulk transition temperature. We see then that, in a film made of the
same material, the critical temperature is diminished by a quantity
proportional to the inverse of its thickness. Also, we see that there is a
minimal film thickness $L_{c}^{(0)}$ below which superconductive is
suppressed, which is given by (identifying the Ginzburg-Landau parameter $ 
\beta =\lambda \mu $) 
\begin{equation}
L_{c}^{(0)}=\frac{3\gamma \beta }{2\sqrt{2}\pi \alpha T_{0}}.
\label{supressao}
\end{equation}
Such a linear dependence of $T_{c}$ with the inverse of the film thickness
has been found experimentally in materials containing transition metals, for
example, in Nb \cite{Nb1,Nb2,Nb3} and in W-Re alloys \cite{W-Re}; for these
cases, it has been explained in terms of proximity, localization and
Coulomb-interaction effects. Notice that our result does not depend on
microscopic details of the material involved nor accounts for the influence
of manufacturing aspects, like the kind of substrate on which the film is
deposited. In other words, the linear decreasing of $T_{c}$ as the film
thickness is diminished emerges solely as a topological effect of the
compactification of the Ginzburg-Landau model in one direction; other
aspects, which may influence the transition temperature of the film, will
show up experimentally as deviations of this linear behavior.

Here, in a field theoretical framework, we have shown that quantum
corrections to the mass in the Ginzburg-Landau model compactified in one of
the spatial dimensions, in one loop-order, leads to the linear dependence of 
$T_{c}$ with the inverse of the thickness for a film, superconductivity
being suppressed at a minimum critical thickness $L_{c}^{(0)}$. One expects,
however, that the inclusion of the $L$-dependence of the coupling constant
in first-order may lead to a small correction to the linear behavior
obtained. Our treatment can also be extended to consider external magnetic
fields, but these issues will be discussed elsewhere.

This paper was partially supported by the Brazilian agencies CNPq and
FAPERJ. One of us (A.P.C.M.) is grateful for kind hospitality to Instituto
de Fisica, UFBA (Brazil), where part of this work has been done.

\end{document}